\documentclass{nature}

\usepackage{amsmath}    % need for subequations
\usepackage{graphicx}   % for figures
\usepackage{color}
\usepackage[10pt]{moresize}
\usepackage{graphics,graphicx}
\usepackage{amsfonts}
\usepackage{amssymb}
\usepackage{amscd}
\usepackage{amsmath}
\usepackage{enumerate}
\usepackage{epsfig}
\usepackage{subfigure}

\let\oldcaption\caption
\renewcommand{\caption}{\sffamily \oldcaption}
\begin{document}

\title{Memory effect of the online user preference}

\author{Lei~Hou$^1$, Xue~Pan$^1$, Qiang~Guo$^1$, Jian-Guo~Liu$^{1,\ast}$}

\maketitle

\begin{affiliations}
\item
Research Center of Complex Systems Science, University of Shanghai for Science and Technology, Shanghai 200093, P.~R.~China\\
$^\ast$ To whom correspondence should be addressed. E-mail: liujg004@ustc.edu.cn
\end{affiliations}

\baselineskip24pt

\maketitle

\begin{abstract}
The mechanism of the online user preference evolution is of great significance for understanding the online user behaviors and improving the quality of online services. Since users are allowed to rate on objects in many online systems, ratings can well reflect the users' preference. With two benchmark datasets from online systems, we uncover the memory effect in users' selecting behavior which is the sequence of qualities of selected objects and the rating behavior which is the sequence of ratings delivered by each user. Furthermore, the memory duration is presented to describe the length of a memory, which exhibits the power-law distribution, i.e., the probability of the occurring of long-duration memory is much higher than that of the random case which follows the exponential distribution. We present a preference model in which a Markovian process is utilized to describe the users' selecting behavior, and the rating behavior depends on the selecting behavior. With only one parameter for each of the user's selecting and rating behavior, the preference model could regenerate any duration distribution ranging from the power-law form (strong memory) to the exponential form (weak memory).
\end{abstract}

Collective behaviors have been investigated for decades and have been proved to be regular more than random. Especially in recent years, thanks to the information technology and computer network, physicists and sociologists have uncovered many remarkable statistical properties and patterns of collective behaviors with massive data. The human mobility has been found following reproducible and predictable patterns \cite{mobility1, mobility2, mobility3}. The communication pattern shows the bursty nature, exhibiting heavy-tailed distribution for the inter-event time \cite{communication1, communication2, communication3,communication4}. Many models have been proposed to describe the patterns and fundamental mechanisms of those collective behaviors, such as the task-based queuing model \cite{communication1, queuing2, queuing3} and the interest-driven model \cite{interest-driven} describing the origin of the bursty nature, the radiation model \cite{radiation} describing the migration and mobility patterns. Numerous scientists are going further and deeper on the road of understanding the collective behavior's mechanism.

Among these collective behaviors, the online user behavior got more and more attention \cite{Onnela, anchoring-bias, Markovian1, Markovian2, Havlin1, Havlin2, interest-dynamic, activity, online1, online2, online3} with the rapid development of Internet. Despite the bursty nature of inter-event time, efforts also have been paid to investigate the behavior itself, such as the social influence of the decision making when installing online applications \cite{Onnela, Hu} and the anchoring bias of online rating \cite{anchoring-bias}. Although the collective behavior's inter-event time exhibits high-burstiness-low-memory property \cite{burstiness-memory}, evidences of memory effect of the online user behavior itself have been brought out. The Markovian process is widely used to model the users' web browsing patterns \cite{Markovian1, Markovian2}, assuming that, the user's next action depends only on his/her current action. Actually, we can consider this kind of Markovian type patterns having short memories because there are correlations between every two continuous actions of a user. Furthermore, using the method of detrended fluctuation analysis, Rybski {\it et al.} \cite{Havlin1, Havlin2} found the long-term memory of users' online communicating frequency.

Nowadays, the Internet does lots of favor for our daily life. The most frequent action of our online behavior is selecting, which is a reflection of our online preference, such as selecting commodities, music or movies. Thus, the memory effect representing the predictability of users' online preference is of great significance for developing the recommendation systems \cite{recommender1, recommender2, recommender3} and providing better online service. But the question is, what is the mechanism governing our online preference? A related work refers to the recent study of the users' commodity-browsing behavior \cite{interest-dynamic}. Zhao {\it et al.} found that, when browsing commodities, the intervals of users staying in a single catalog present a power-law distribution, which indicates the users tend to continuously browse similar commodities. On the other hand, users may return to historical catalogs after some specific intervals. This kind of online commodity-browsing behavior has memory effect and is quite predictable which is similar to the human mobility behavior \cite{mobility3}. Despite the memory effect of transferring from a catalog to another, are there any correlations between every two continuous actions or how do the users feel like their selections? Our goal here is to uncover and model the local correlations of the dynamics of online user preference when selecting and evaluating movies. As many online systems allow users to deliver ratings on objects which could largely reflect how the user feels about the objects \cite{LIYANG}, we adopt another method besides the catalogs to evaluate objects in a more detailed way.

{\bf Data.} - Two datasets, namely {\it MovieLens} and {\it Amazon}, are investigated in this paper. The MovieLens data consists of 698054 ratings delivered by 5547 users on 5850 movies during 1686 days. The Amazon data consists of 1406147 ratings of 624271 users rated on 86087 items. We uniformly call both the movies in the MovieLens data and the items in the Amazon Data as {\it objects} in this paper. Each rating in both datasets is an integer ranging from 1 to 5 reflecting how the user feels like the objects' worth, taste and so on. The higher a rating is, the better the user evaluates the object. Then, we could also define the {\bf quality} of an object as the average value over all ratings the object got in the whole dataset. Thus, we can get two messages from the user's behavior: the quality of the object he/she selects and the value of the rating. Those two messages can be regarded as the user's selecting behavior and rating behavior respectively. Consequently, we investigate the {\bf Selecting Series (SS)} which is the sequence of the object quality of each user's choice (a typical example is shown in Fig. 1 (d)) and the {\bf Rating Series (RS)} which is the sequence of each user's ratings (a typical example is shown in Fig. 1 (a)). These two series are sequences of users' behavior in order of time. We will firstly uncover and model the self-correlation of users' SS and RS ignoring the exact inter-event time, and then discuss the inter-event time's effect on the self-correlation. It should be noted that, while the MovieLens data is ordered by seconds, the Amazon data is ordered by days. Thus, a few records of Amazon data involved in the situation that, at the same day, a user selects several objects which cannot be ordered by time according to the data. Against this kind of situations, we arrange the records of a specific user occurring on the same day in random order. Furthermore, the {\it activity level} of each user, i.e. the number of objects the user has selected, is different. We denote the length with $L$ and in order to ensure the accuracy of the results, we only take users with $L\geq100$ into consideration.

{\bf Methods} - To evaluate the memory effect of the user's SS and RS, we use the method of Correlation Coefficient which is also used by Goh and Barab\'{a}si \cite{burstiness-memory}. The memory $M$ is defined as
\begin{eqnarray} \label{eq.1}
M=\frac{1}{L-1}\sum_{i=1}^{L-1}\frac{(r_i-m_1)(r_{i+1}-m_2)}{\sigma_{1}\sigma_{2}},
\end{eqnarray}
where $r_i$ is the $i$th value in the user's series, and $m_1$($m_2$) and $\sigma_{1}$($\sigma_{2}$) are the mean and standard deviation of the sample series $\{r_1, r_2, \cdots, r_{k-1}\}$ $(\{r_2, r_3, \cdots, r_k\})$ respectively. Actually, the correlation coefficient method could measure the correlation between two continuous values in a single sequence. Thus, the correlation coefficient could well describe the local memory effect. With this definition, the memory $M$ has a value in the range $(-1, 1)$ and is positive when a high(low) value in the series tends to be followed by a high(low) one, and it is negative when a high(low) value in the series tends to be followed by a low(high) one. Note that, the present paper aims to study the users' memory effect on the values of ratings and qualities, not the inter-event times \cite{burstiness-memory} and the object catalogs \cite{interest-dynamic}.

There are also other methods evaluating the memory effect of series, for example, by using detrended fluctuation analysis \cite{Havlin1, Havlin2, DFA} to detect the Hurst exponent \cite{HURST} of the series. However, the length of series should reach $10^4$ to guarantee the estimation of Hurst exponent \cite{scale} and furthermore, the Hurst exponent is to describe the long-term memory. In this paper, the length of series is of scale $10^2$, and we mainly focus on the local correlation, thus, we didn't use the method of Hurst exponent.

Using the correlation coefficient method, the user shown in Fig.1 has memory effect $M_{RS}=0.216$ and $M_{SS}=0.471$ for RS and SS respectively. The positive value of this user's memory effect $M_{RS}$ and $M_{SS}$ means that, when the user selected a high(low)-quality object, he/she would continuously select high(low)-quality objects, and further, when he/she delivered a high(low) rating, he/she would continuously deliver high(low) ratings. In other words, a memory of this user's preference may last for several rounds of actions. But the question is, how long this kind of memory could last? We define in this paper that, when the value in a user's series changing from less(greater)-than-mean to greater(less)-than mean, the user's current memory ends and his/her next memory begins. Thus, a user may has many memories, and we further define the {\bf memory duration $\lambda$} as each memory's length, i.e., the number of continuous qualities or ratings which are greater(less) than the mean value of the user's SS or RS, as shown in the Fig. 1 (b) and (e). To give an example, for a specific user, one of his/her memory's duration is $\lambda=4$ for RS means that, he/she continuously delivered 4 ratings which are all greater than the mean value of his/her RS. Note that, the summation of the memory durations over all of a user's memory equals to the length of the behavior sequence $\sum_{i=1}^{k_u}\lambda_i=L$, where $\lambda_i$ is the duration of the user's $i$th memory and $k_u$ is the number of user $u$'s memories. Counting every memory duration, we can calculate the probability $p(\lambda)$ of a memory with duration $\lambda$, and the duration distributions for RS and SS of the typical user are reported in Fig. 1 (c) and (f).

{\bf Empirical Results. - } For each user, we calculate the memory $M$ of his/her SS and RS respectively according to Eq. (1). The distribution of memory $M$ is shown in Fig. 2 together with that of the null model. The null model is the case in which we shuffle each user's series into random order to remove the temporal behavior, so that the users' selecting and rating patterns no longer exist. The mean value of the memory $M$ of RS and SS is 0.19 and 0.36 respectively for the MovieLens data and is 0.18 and 0.17 respectively for the Amazon data. This result indicates the existence of memory effect of users' both rating and selecting behavior. On the other hand, the memory $M$ of the null model has a mean value of 0. We can conclude from the comparisons between the null model and the empirical data that, the memory effect of users' selecting and rating behavior comes from their own behavior patterns not the random mechanism.

As shown in Fig.1 (c) and (f), the duration distributions $p(\lambda)$ approximately exhibit power-law decays on the individual level. That is to say, there are probabilities occurring memories with very long duration. For the typical user in Fig.1, whose activity is $L=464$, the memories may last for about 30 rounds of actions for both rating and selecting behavior. Actually, on the individual level, users with different activity levels $L$ all approximately have the power-law duration distribution (Fig. S1). On the collective level, the duration distribution $p(\lambda)$ also exhibits the power-law form, as shown in Fig. 3. Thinking of the totally random case, the probability of a user selecting a high(low)-quality object at next time is 0.5. Thus, the theoretical duration ditribution should be $p(\lambda)=0.5^{\lambda}$ which has the exponential form for the totally random case. As shown in Fig. 3, the duration distributions of the shuffled series for both the MovieLens and the Amazon data are similar to that of the totally random case, following the exponential form. That is to say, once the temporal pattern is removed, users' online behaviors would perform high randomness. On the other hand, the empirical results are very different with the totally random case, with power-law form duration distribution which suggests that, the sequence of users' selecting and rating behavior is self-correlated rather than random. Note that, although the probability of the memory having a long duration is small (for example, $p(\lambda>10)=0.042$ for MovieLens' SS), the amount of user's actions involved in a long duration memory is apparent. About 28.3\% of the users' actions involved in memories with duration larger than 10, which are quite long memories (Fig. S2).

{\bf Preference Model.} - Results from the empirical data reported the memory effect of users' selecting and rating behaviors. Modelling the mechanism is crucial for understanding the dynamic of users' online preference. Hereinafter we model the users' selecting and rating behaviors in two steps.

1) Selecting behavior. Suppose a user's next selection depends on the current selection, which results in the memory effect of the user's SS. Thus, we use the Markovian process to model the selecting behavior mechanism. The Markovian process in users' selecting behavior is the process in which the users' current selection with quality $q_i$ transfers to the next selection with quality $q_{i+1}$. The empirical statistics of the bias $\delta_{SS}=q_{i+1}-q_i$ are shown to follow Gaussian forms with expectations $\mu_{SS}=0$ (Fig. S3). Furthermore, the standard deviations are $\sigma_{SS}^{\rm mov}=0.563$ and $\sigma_{SS}^{\rm ama}=0.701$ for the MovieLens and the Amazon data respectively. Thus, the transition probability distribution of the Markovian process should be a Gaussian distribution with expectation $\mu=q_i$ and standard deviation $\sigma_{SS}$. Then, the quality of the user's next selection $q_{i+1}$ comes from this Gaussian distribution, i.e. the probability of the next object's quality being $q_{i+1}$ is given by
\begin{equation}
f(q_{i+1})=\frac{1}{\sigma_{SS}\sqrt{2\pi}}{\rm exp}(-\frac{(q_{i+1}-q_i)^2}{2\sigma_{SS}^2}).
\end{equation}

2) Rating behavior. We assume that, while the selecting behavior is self-correlated between current and next actions, the memory effect of the rating behavior origins in the selecting behavior. The value of rating $r_i$ only depends on the quality $q_i$ of the selected object. Similar with the selecting bias, the rating bias $\delta_{RS}=r_i-q_i$ also exhibit Gaussian distributions for both the MovieLens and the Amazon data (Fig. S3). The expectations are $\mu_{RS}^{\rm mov}=0$ and $\mu_{RS}^{\rm ama}=0.056$ and the standard deviations are $\sigma_{RS}^{\rm mov}=0.961$ and $\sigma_{RS}^{\rm ama}=0.964$ for the MovieLens and the Amazon data respectively. Thus, we can also give the probability of the rating the user deliver on a object with quality $q_i$ being $r_i$ as
\begin{equation}
f(r_i)=\frac{1}{\sigma_{RS}\sqrt{2\pi}}{\rm exp}(-\frac{(r_i-q_i)^2}{2\sigma_{RS}^2}).
\end{equation}

The quality of an object could be a float value ranging from 1 to 5, and the rating $r_i$ is an integer. In those two stochastic processes, we continuously get random values until the value locates in the range [1, 5]. In addition, for the RS, we round the value into integer. When generating series, with respect to the empirical data, we generate series with length $L_i$ for each of the user $i$, i.e. we remain the number of users and the activity levels of each of them. As to the initial condition $q_0$, we give a mid-value of the range of quality $[1, 5]$, i.e., $q_0=3$.

There are two free parameters left in the preference model, that is the standard deviations of the two Gaussian distributions $\sigma_{SS}$ and $\sigma_{RS}$. Despite that the statistics have shown the empirical value of the standard deviation $\sigma_{SS}$ and $\sigma_{RS}$ for the MovieLens and Amazon data (Fig. S3), those two parameters could control the distributions of the memory duration $p(\lambda)$. We simulate with different values of $\sigma_{SS}$ and $\sigma_{RS}$ for the MovieLens system, and the distributions of the memory duration $p(\lambda)$ is shown in Fig. 4. When those two parameters are small, the memory duration of both SS and RS exhibit a well power-law distribution which is similar to the empirical pattern. As the parameters increase, the distributions gradually change from the power-law form to the exponential form and become more and more similar with the totally random case. The reason lies in the fact that, when the standard deviation is small, a burst jump of the values has little chance to occur, i.e. the memory effect is stronger. On the other hand, if the standard deviation is large enough, the Gaussian distribution could be approximately considered as the uniform distribution which leads to the totally random case in which $p(\lambda)=0.5^{\lambda}$. Results of Fig. 4 indicate that, the standard deviations of the SS's and RS's bias indeed could control the distribution of the memory duration. When simulating with the empirical values of $\sigma_{SS}$ and $\sigma_{RS}$ for the MovieLens and Amazon data respectively, the memory duration distributions of the simulation consist with that of the empirical data (Fig. S4). It means that, the preference model can well describe the emergence of heavy-tailed distribution in the memory duration. But on the other hand, the preference model has deviation from the empirical data in the distributions of memory $M$ for users' both SS and RS.  The comparison between the empirical data and the preference model indicates that, users' selecting behavior is approximately a Markovian process and that the rating behavior mainly depends on the selecting behavior. Briefly speaking, when a user selects an object with quality $q_i$ at this time, the quality of the next object $q_{i+1}$ is probably around $q_i$, and the user further tend to deliver a rating which is also around the quality of the object. In addition, the memory effect of users' rating behavior origins in the selecting behavior.

We can reproduce any duration distributions from power-law to exponential form with just one parameter for each of the selecting and rating behavior. The standard deviation could largely explain the users' memory effect and the power-law decay of duration distribution. However, as users with different activity levels have in general different properties \cite{activity, C4, Ceiling, Newman}, the parameter may be various from users to users. Actually, the activity level indeed affects the memory effect (Fig. S5). In the MovieLens system, high-activity users generally have stronger memory effect and lower deviation parameter $\sigma_{SS}$ and $\sigma_{RS}$. But in the Amazon system, the activity level is uncorrelated with the users selecting behavior. For the rating behavior, high activity level would approximately leads to strong memory effect and small deviation parameter $\sigma_{RS}$. Overall, we just need to fit different parameters $\sigma_{SS}$ and $\sigma_{RS}$ to describe and reproduce different users' selecting and rating behavior.

{\bf Effect of the Inter-event Time. - } Numerous investigations proved that, the heavy-tailed distribution of inter-event time was one of the most important properties of collective behavior \cite{communication1, interest-driven, activity}. As the present paper aims to study the correlations between online users' current and next actions, the inter-event time between those two actions may has important influence. As Fig. 5 (a) and (d) show, the distributions of inter-event time for both MovieLens and Amazon system exhibit power-law forms which have been observed in many classical researches. For a user whose activity level is $L$, there would be $L-1$ inter-event times $\tau$. After averaging each user's inter-event times, we show the correlation between user's average inter-event time $\langle\tau\rangle$ and their memory $M$ as shown in subplot (b) and (e). In MovieLens dataset, users who have long average inter-event time generally have weak memory effect and those whose average inter-event times are short have strong memory effect. It is easy to understand that, while there would be a strong correlation between two actions, if one occurred immediately after another, the correlation would be very weak if two actions have very long inter-event time. However, for Amazon system, the users' memory effect is approximately uncorrelated with their average inter-event time. To uncover the time effect on the preference model, we calculate the conditional standard deviation $\sigma_{SS}(\tau)$. Each pair of selecting behaviors with inter-event time $\tau$ has bias $\delta_{SS}(\tau)$, and $\sigma_{SS}(\tau)$ is the standard deviation of $\delta_{SS}(\tau)$. Figure 5 (c) and (f) shows the correlation between $\sigma_{SS}(\tau)$ and $\tau$. It is surprising to find that, the MovieLens and Amazon system has different reactions toward the inter-event time in the parameter $\sigma_{SS}(\tau)$. As we reported, $\sigma_{SS}(\tau)$ could reflect the strength of the memory effect. Subplot (c) shows that, the longer the inter-event time is, the weaker the memory effect would be in the MovieLens system. But for the Amazon system which is shown in subplot (f), the situation is totally different that, the longer the inter-event time is, the stronger the memory effect would be. The difference of the inter-event time's effect between those two datasets may lies in the fact that, MovieLens is a system in which users watch and rate movies, but Amazon is a system in which users buy items. In the Amazon dataset, we observed the phenomenon that users might select many same objects. Those repeated objects could be consumables that, users may buy a new one after specific times intervals. Furthermore, the evaluation of a user to a very object is in general similar. This may be the reason that, long inter-event time brings strong memory effect.

{\bf Discussion.} - Considering the significance of online user preference for the understanding of collective behavior pattern and the developing of online systems, we investigated the memory effect of users' selecting and rating behavior with the method of Correlation Coefficient. The mean value of memories $M$ for SS and RS of the empirical data is 0.36 and 0.19 for the MovieLens dataset and 0.17 and 0.18 for the Amazon dataset respectively, indicating that, the complex online user preference dynamics have memory effect. Furthermore, we found the distribution of the memory duration, which was used to describe the memory's length, exhibiting scaling law with heavy-tailed power-law form. To model the pattern and the fundamental mechanism of online user preference, we utilized a Morckovian process to model the selecting behavior and supposed the rating behavior totally depending on the selecting behavior. The distribution of the memory duration of the preference model coincided with the empirical data. Just one parameter for each of the RS and SS is needed to reproduce any duration distribution ranging from the power-law to exponential form.

Results in this paper indicated that, the Markovian process could largely explain the memory effect of users' online selecting behavior, and the memory effect of the rating behavior origins in the selecting behavior. In a recent study \cite{anchoring-bias}, Yang {\it et al.} found a correlation between the quality of the former selecting $q_{i-1}$ and the current rating $r_i$, that, when $q_{i-1}$ is very high(low), $r_i$ would also be high(low). This anchoring bias phenomena can be explained by the present paper. Actually, when $q_{i-1}$ is high(low), the quality of the current selection $q_i$ would also be high(low) according to the memory effect. Furthermore, as the rating behavior largely depends on the selecting behavior, the current rating $r_i$ would consequently also be high(low).

However, we cannot ensure the effect of inter-event time and the activity level. As active users have in general shorter inter-event times, we could not know whether the activity level or the inter-event time or both of them lead to the heterogenous memory effect $M$ shown in Fig. 5 (b) and (e) and Fig. S5 (a) and (d). In addition, the MovieLens and the Amazon datasets exhibit differences in some results such as the activity's and inter-event time's effect on the memory and the preference model. What caused the different patterns for those two systems is an important question. As we have introduced that, the MovieLens is a system in which users watch and rate movies, but the Amazon is a system in which users buy items. While watching movies is an entertainment behavior, buying items would cost money. Users' selecting and rating behaviors are based on different purposes and considerations in these two systems. Thus, the question is, do the patterns vary for different types of behaviors?

Absolutely, our effort is still far from totally understanding the online user behavior patterns. When modeling the users' selecting behavior, we utilized the Morckovian process in which the user's next selection depends only on his/her current selection. We didn't consider the possibility that, the next selection might be correlated with not only the current one but also former several ones. This is a kind of very short and local correlation, but whether there is a long-term correlation and whether it is possible for a user to repeat some selecting behavior fragments are still open questions. In addition, many other factors could affect user's selecting and rating behavior, such as the social influence \cite{Onnela, Hu} or the recommendation list. Another problem is, the quality (average rating) is still not enough to describe all the aspects of an object. It's reasonable to consider other aspects seriously such as the popularity.

\begin{addendum}

\item  This work is partially supported by NSFC (71371125, 61374177 and 71171136), Shanghai Leading Academic Discipline Project (Systems Science) (No. XTKX2012) and MOE Project of Humanities and Social Science (13YJA630023, 14ZR1427800).

\item[Author Contributions] L.H. and J.G.L. conceived the idea. L.H. and X.P. performed the calculation. L.H., X.P., Q.G. and J.G.L. discussed the results and wrote the manuscript. Correspondence and requests for materials should be addressed to J.G.L. (liujg004@ustc.edu.cn).

\item[Competing Interests] The authors declare that they have no competing financial interests.

\end{addendum}

\begin{figure}%[!htb]
\begin{center}
\scalebox{.8}[.8]{\includegraphics{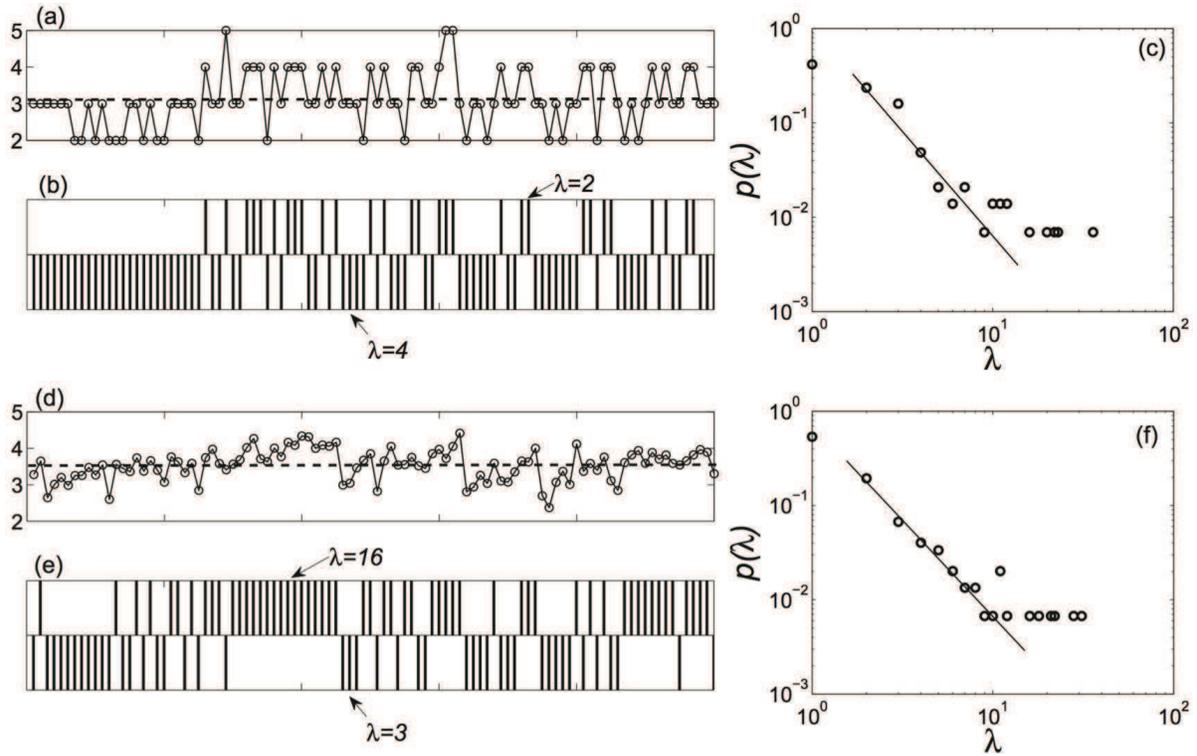}}
\caption{A typical user's RS, SS and duration distribution $p(\lambda)$ from the MovieLens data. This user's activity level is $L=464$, and his/her memory effect is $M_{RS}=0.216$ and $M_{SS}=0.471$ for RS and SS respectively. (a), (d) A part of the user's whole RS and SS respectively. While the values in RS are integers, SS consists of float values ranging from 1 to 5. Regarding those values which is greater than the mean value as positive bars and those which is less than the mean value as negative bars, one can get the subplot (b) and (e) for RS and SS respectively. While every user may have many memories, we define in this paper that, a memory ends only when the bar in his/her series changing from positive (negative) to negative (positive). Then, the memory's duration $\lambda$ is the length of the series involved in that memory, as typical marks shown in subplot (b) and (e). (c), (f) The memory duration distribution $p(\lambda)$ of the typical user for RS and SS respectively.}
\end{center}
\end{figure}

\begin{figure}%[!htb]
\begin{center}
\scalebox{.8}[.8]{\includegraphics{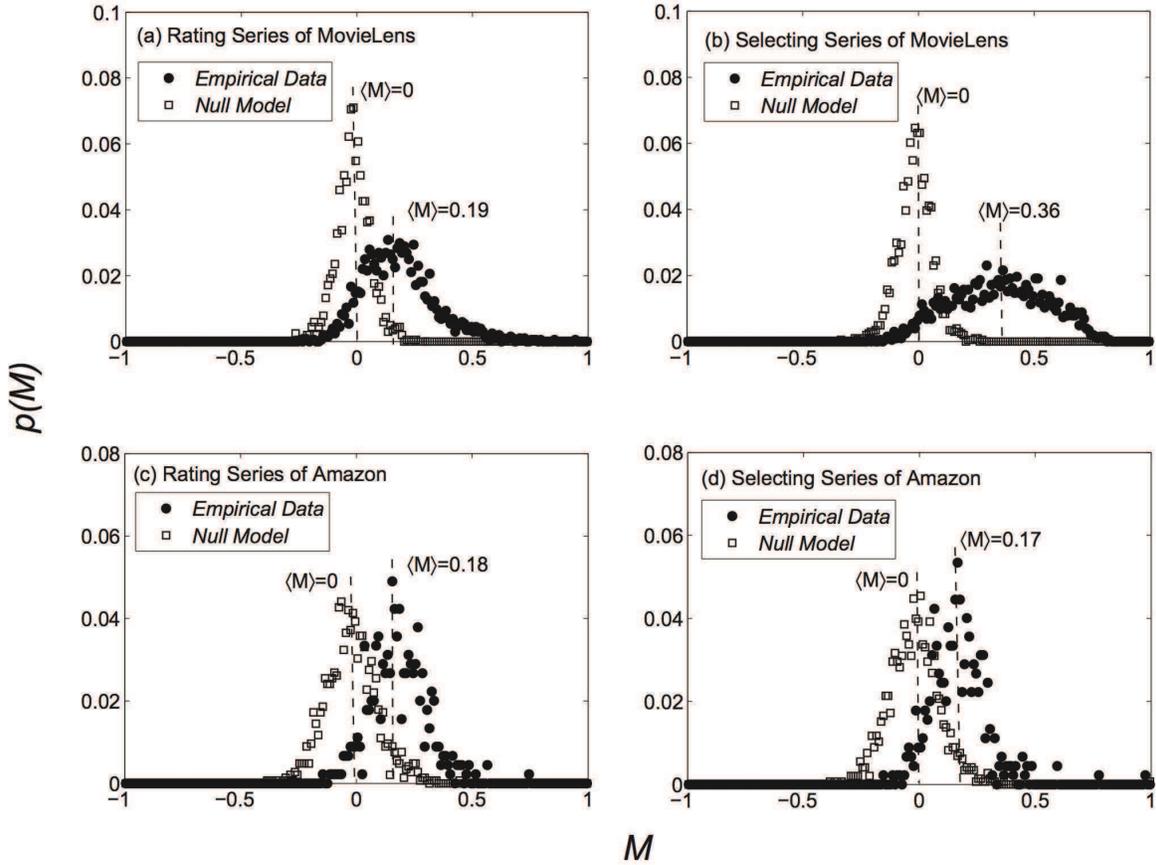}}
\caption{The distributions of memory $M$ for the MovieLens and the Amazon datasets. Subplots (a) and (b) show the distributions of RS and SS for the MovieLens data respectively, in which, the memory $\langle M_{RS}\rangle=0.19$ and $\langle M_{SS}\rangle=0.36$. Subplots (c) and (d) show the distributions of RS and SS for the Amazon data respectively, in which, the memory $\langle M_{RS}\rangle=0.18$ and $\langle M_{SS}\rangle=0.17$. As to the null model, the memory $M$ distribute in a narrow range and the mean value $\langle M\rangle=0$. Comparisons between the empirical data and the null model report the existence of the memory effect.}
\end{center}
\end{figure}

\begin{figure}[!htb]
\begin{center}
\scalebox{0.45}[0.45]{\includegraphics{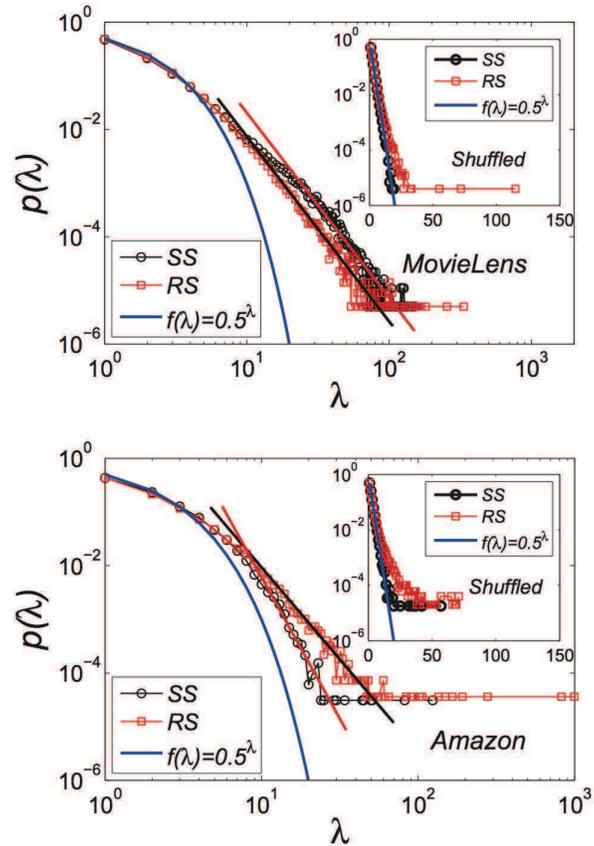}}
\caption{(Color online) Memory duration distributions $p(\lambda)$ of the empirical data, the null model, and the totally random case for the MovieLens data (above) and the Amazon data (bottom). While duration $\lambda$ of the empirical data follows the power-law distribution, that of the null model approximately follows the exponential distribution, which is very similar with the totally random case in which $p(\lambda)=0.5^{\lambda}$. Note that, the shuffled RS partly deviates from the totally random case in both datasets. The reason lies in the fact that, the values in RS are integers ranging from 1 to 5, with only 5 options, and some users tend to deliver the same ratings to whatever he/she selected without consideration. Thus, some users' RS may consist of large amount of same values which would lead to the continuous same values for the shuffled RS.}
\end{center}
\end{figure}

\begin{figure}[!htb]
\begin{center}
\scalebox{.45}[.45]{\includegraphics{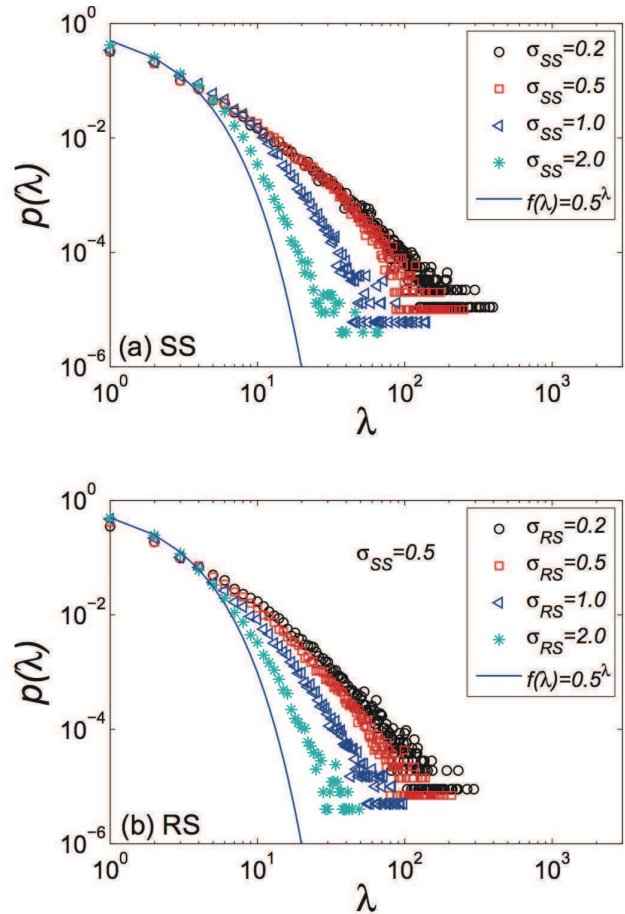}}
\caption{(Color online) Memory duration distributions $p(\lambda)$ of SS and RS from the simulations of preference model with different values of parameter $\sigma_{SS}$ and $\sigma_{RS}$ respectively. When simulate the preference model for the RS, we arrange the standard deviation of SS's bias as $\sigma_{SS}=0.5$. As the results show, for both SS and RS, the smaller the standard deviation is, the heavier the distribution's tail would be. On the other hand, as the standard deviation increases, the distribution gradually changes from the power-law form to the exponential form, and is more and more similar to the distribution of the totally random case.}
\end{center}
\end{figure}

\begin{figure}[!htb]
\begin{center}
\scalebox{.8}[.8]{\includegraphics{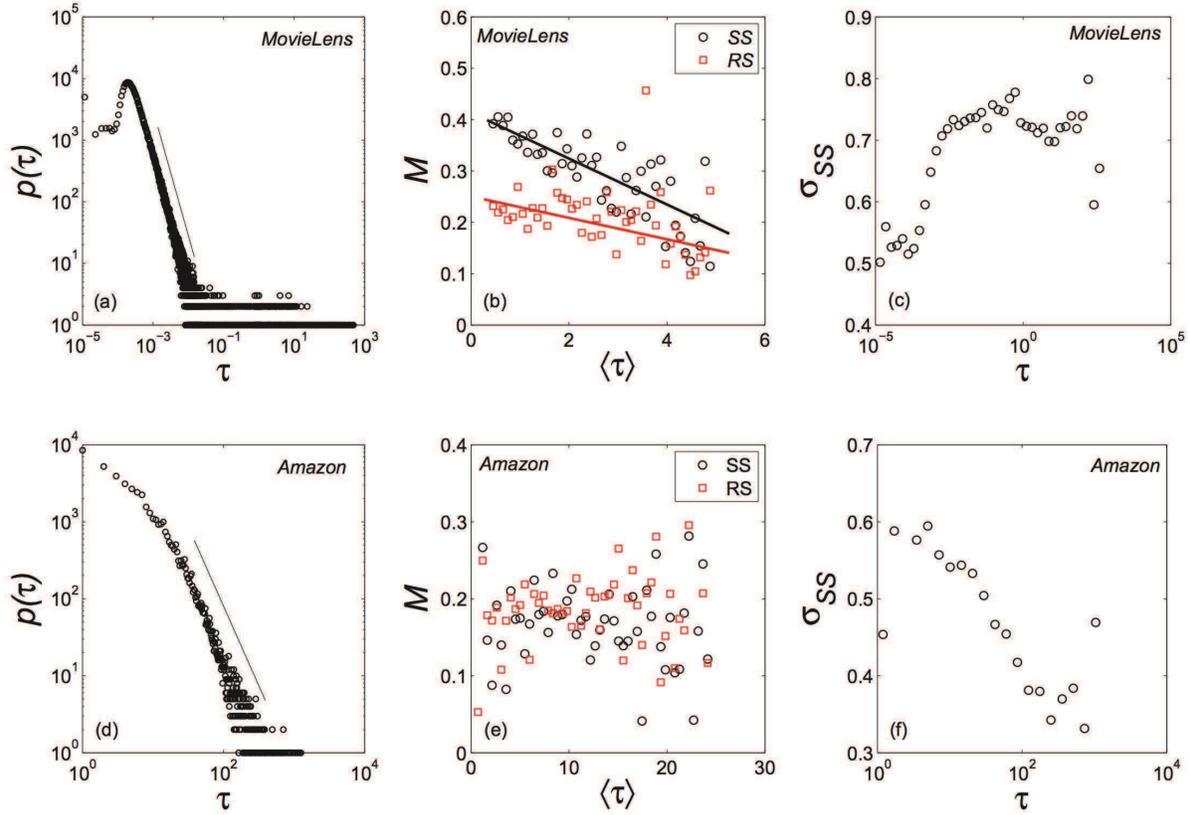}}
\caption{(Color online) Inter-event time distribution $p(\tau)$ and its effects on the memory effect $M$ for the MovieLens and the Amazon data. Subplots (a) and (d) report the inter-event time distributions on the collective level. As has also been reported in many classical researches, the inter-event time exhibits heavy-tailed power-law distributions. Subplots (b) and (e) show the correlation between memory effect $M$ and the user's average inter-event time $\langle\tau\rangle$. Subplots (c) and (f) show the correlation between the deviation parameter $\sigma_{SS}$ and inter-event time $\tau$.}
\end{center}
\end{figure}

%%%%%%%%%%%%%%%%%%%%%%%%%%%%%%%%%%%%%%%%%%%%%%%%%%%%%%%%%%%%%%%%%%%
%\ifthenelse{\boolean{SubmittedVersion}}{\processdelayedfloats}{\cleardoublepage}
\end{document}